\documentclass[epj-spec]{svjour}

\usepackage{amsmath}
\usepackage{graphicx}
\usepackage{color}

\newcommand{\figcopy}{[Taken from P.~Sj\"{o}lund, S.~J.~H. Petra,
  C.~M. Dion, S.~Jonsell, M.~Nyl\'{e}n, L.~Sanchez-Palencia, and
  A.~Kastberg, Phys. Rev. Lett.  \textbf{96}, 190602 (2006). Copyright
  (2006) by the American Physical Society.]} 

\newcommand{\Dv}{\ensuremath{{D_\mathrm{v}}}}
\newcommand{\vect}[1]{\mathbf{#1}}

\begin{document}

\title{Controllable 3D atomic Brownian motor in optical lattices}

\author{Claude M. Dion\inst{1}\fnmsep\thanks{\email{claude.dion@tp.umu.se}} 
  \and Peder Sj\"{o}lund\inst{1} \and Stefan J. H. Petra\inst{1} 
  \and Svante Jonsell\inst{1}\fnmsep\thanks{Current address:
    Department of Physics, University of Wales Swansea, Singleton
    Park, Swansea SA2 8PP, United Kingdom} 
  \and Mats Nyl\'{e}n\inst{1} \and Laurent
  Sanchez-Palencia\inst{2} \and Anders Kastberg\inst{1}}

\institute{Department of Physics, Ume{\aa} University, SE-90187 Ume{\aa},
  Sweden 
  \and 
  Laboratoire Charles Fabry de l'Institut d'Optique, CNRS,
  Univ. Paris-Sud, Campus Polytechnique, RD-128, F-91127
  Palaiseau cedex, France}

\abstract{We study a Brownian motor, based on cold atoms in optical
  lattices, where atomic motion can be induced in a controlled manner
  in an arbitrary direction, by rectification of isotropic random
  fluctuations.  In contrast with ratchet mechanisms, our Brownian
  motor operates in a potential that is spatially and temporally
  symmetric, in apparent contradiction to the Curie
  principle. Simulations, based on the Fokker-Planck equation, allow
  us to gain knowledge on the qualitative behaviour of our Brownian
  motor. Studies of Brownian motors, and in particular ones with
  unique control properties, are of fundamental interest because of
  the role they play in protein motors and their potential
  applications in nanotechnology. In particular, our system opens the
  way to the study of quantum Brownian motors.}

\maketitle

\section{Introduction}

Brownian motors are devices capable of converting the energy of the
random, isotropic motion of Brownian particles into useful work, for
instance driving the particles into a directed motion, without any
macroscopic force \cite{brown:haenggi05a,brown:reimann02a}.  This
possibility is not trivial in the face of fundamental symmetry and
thermodynamic laws.  Indeed, realising a Brownian motor requires that
the system must be (i) asymmetric and (ii) brought out of
thermodynamic equilibrium.  On one hand, the need for asymmetry to
extract directed motion out of random fluctuations is intuitive and is
underpinned by the Curie principle, which states that asymmetric
dynamics cannot emerge in a system possessing both spatial and
temporal symmetries~\cite{brown:curie94}.  On the other hand, the need
for working out of equilibrium comes from the second law of
thermodynamics, which states that the total entropy always increases.
Surprisingly enough, these two requirements are generally sufficient
for realising a Brownian motor, although no rigorous proof is
available so far \cite{brown:reimann02a}.  In his lectures of physics,
Richard Feynman describes a seminal ratchet mechanism able to rectify
noise~\cite{ratchet:feynman63}, based on an original idea of
Smoluchowski~\cite{ratchet:smoluchowski12}.  Up to now, essentially
all suggestions and tentative demonstrations of ratchet effects and
Brownian motors rely on that archetype principle, based on the
application of a force, asymmetric either in space or in time, albeit
one whose macroscopic average vanishes.  However, it was suggested in
ref.~\cite{brown:sanchez-palencia04} that a ratchet effect can be
induced in spatially and temporally symmetric potentials, provided
that asymmetric jumps occur between potentials that are spatially
shifted.  It should be noted also that previous realisations of
Brownian motors typically inherently lack the possibility of inducing
motion in any direction in three dimensions and are difficult to
control, as the asymmetry is built into the system.

Interestingly, systems of cold atoms in dissipative optical lattices
offer plenty of possibilities to investigate standard problems of
statistical physics with an unprecedented
accuracy~\cite{lat:grynberg01}. Indeed, these have proved to be highly
controllable and versatile systems~\cite{lat:grynberg01,lat:jessen96}.
Hence, dissipative optical lattices have been used recently to study
several effects, such as mechanical
bistability~\cite{lat:grynberg00b,lat:visser00}, spatial diffusion in
random or quasi-periodic
structures~\cite{lat:horak98,lat:grynberg00a,lat:guidoni97,lat:guidoni99b},
and stochastic
resonance~\cite{lat:sanchez-palencia02b,lat:schiavoni02,%
  lat:sanchez-palencia03c}. Ratchet effects have also been
investigated in dissipative optical lattices with either a
spatial~\cite{lat:mennerat-robilliard99,lat:robilliard02}, or a
temporal asymmetry~\cite{lat:schiavoni03,lat:jones04,lat:gommers05a,%
  lat:gommers05b}.

In this work, we review our recent experimental realisation of a
Brownian motor based on the model proposed in
ref.~\cite{brown:sanchez-palencia04}, using an ultra-cold gas of atoms
trapped in a stationary dissipative double optical
lattice~\cite{lat:sjolund06}, and provide further analysis of the
rectification mechanism.  Our Brownian motor operates with a mechanism
where the potentials are both temporally and spatially
symmetric. Moreover, in contrast to other rectification mechanisms
reported so far, the motion can be induced in any direction in three
dimensions with a controlled speed. This new type of Brownian motor
opens up possibilities for fundamental studies of noise-induced
directed motion. The underlying principle is very general and is
potentially transferable to molecular motors and to applications in
nano-electronics and chemistry~\cite{brown:sanchez-palencia04}.

We start by presenting in sec.~\ref{sec:BM} the basic working
principle of our Brownian motor.  Then, in sec.~\ref{sec:exp}, we
discuss its experimental realisation with cold atoms in a double
optical lattice.  In sec.~\ref{sec:simul}, we investigate the
qualitative behaviour of the Brownian motor, using classical
simulations based on the Fokker-Planck equation.  Then, we discuss in
sec.~\ref{sec:quantum} how this system could be extended to a regime
relevant for the study of quantum Brownian motors.  Finally, we
summarise our results in sec.~\ref{sec:conclusion}.

\section{Brownian motor with symmetric potentials}
\label{sec:BM}

\begin{figure}[b!]
\centerline{\includegraphics[width=0.45\textwidth]{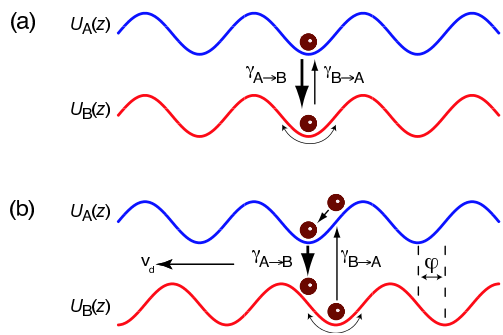}}
\caption{Rectification mechanism. Atoms move in two symmetric
  potentials $U_{\mathrm{A}}(z)$ and $U_{\mathrm{B}}(z)$ that are
  coupled via the asymmetric optical pumping rates $\gamma_{
    \mathrm{A} \rightarrow \mathrm{B}}$ and $\gamma_{ \mathrm{B}
    \rightarrow \mathrm{A}}$ ($\gamma_{ \mathrm{A} \rightarrow
    \mathrm{B}} \gg \gamma_{ \mathrm{B} \rightarrow \mathrm{A}}$). a)
  The potentials are in phase. The transfer from the long lived state
  B, to the transient state A, and back, will not lead to biased
  motion. b) A phase shift $\varphi~(\neq 0,\pi)$ is
  introduced. Spatial diffusion will be strongly facilitated in one
  direction, leading to a drift velocity $v_\mathrm{d}$. \figcopy}
\label{fig:model}
\end{figure}

The basic rectification mechanism demonstrated in this work is
depicted in fig.~\ref{fig:model} (see also
refs.~\cite{brown:sanchez-palencia04,lat:sjolund06}). 
Depending on their internal state, the atoms are subjected to one of
two three-dimensional periodic potentials ($U_\mathrm{A}$ and
$U_\mathrm{B}$, represented in 1D in fig.~\ref{fig:model}) with
identical periods.  In the simplest situation, the potentials are
identical but possibly spatially shifted.  At sufficiently low
temperature, the atoms are trapped in the wells of the potentials.  In
addition, they are assumed to undergo a Brownian motion in each
potential and can be transferred to a neighbouring site via thermal
activation. Ever so often, an atom will be pumped from one internal
state to the other, resulting in random jumps between potentials
$U_\mathrm{A}$ and $U_\mathrm{B}$, with rates $\gamma_\mathrm{A
  \rightarrow B}$ and $\gamma_\mathrm{B \rightarrow A}$.  The
asymmetry that eventually gives rise to controlled rectification is
caused by a pronounced difference in the transfer rates between the
potentials ($\gamma_\mathrm{A \rightarrow B} \neq \gamma_\mathrm{B
  \rightarrow A}$).  In the case where $\gamma_\mathrm{A \rightarrow
  B} > \gamma_\mathrm{B \rightarrow A}$, an atom will spend most of
the time in the long lived state (B), and will oscillate near the
bottom of a trapping site.  At random times, it is pumped to the
transient state (A), from where it returns quickly to state B.  This
excursion may drastically affect the motion of the atoms.

If the two potentials are identical and in phase
[fig.~\ref{fig:model}(a)], this excursion will not affect the dynamics
of the atoms. If the potentials are different, but still in phase, the
change of state may result in a slight heating and increase the
probability for an atom to be transferred to a neighbouring trapping
site.  This diffusion is symmetric. Qualitatively, the same occurs if
the potentials are shifted by half a spatial period of the lattices,
since no direction is favoured.  The situation changes drastically if
the relative phase between the potentials is shifted
[fig.~\ref{fig:model}(b)]. During the time spent in lattice A, the
atom experiences a potential with an incline that depends on the phase
shift. The diffusion is then enhanced in a given direction, and
correspondingly reduced in the opposite direction. While the
potentials are symmetric and stationary, the atoms are propelled in a
specific direction, which can be controlled by modifying the phase
shift.

It is straightforward to extend the principle to three-dimensional
potentials and to obtain a controlled motion in any direction.  This
mechanism works despite the absence of spatial as well as of temporal
asymmetry, in \emph{apparent} contradiction to the Curie
principle~\cite{brown:curie94}. Instead, the combined dynamics, made
up of phases of Hamiltonian motion interrupted by stochastic
dissipative processes, provides the
asymmetry~\cite{brown:sanchez-palencia04}.  In our system, the
directed motion is induced for atoms switching between two
state-dependent periodic potentials that are coupled via optical
pumping. The rectification process emanates from the fact that the
couplings between the two potentials used, via the vacuum field
reservoir, are strongly asymmetric.

\section{Experimental realisation}
\label{sec:exp}


This Brownian motor is realised, in a completely controllable fashion,
using cold caesium atoms in a double optical
lattice~\cite{lat:sjolund06,lat:ellmann03a,lat:ellmann03b,lat:petra06}. The
interaction of an atom with the interference pattern from a number of
laser beams creates a spatially periodic potential (optical lattice),
due to a second-order interaction between an atomic dipole and the
light field~\cite{lat:grynberg01,lat:jessen96}. With the laser
frequency of the optical lattice close to an atomic resonance,
spontaneous emission of photons leads to dissipation. In addition to
the trapping potential, we then have Sisyphus
cooling~\cite{cool:chu98,cool:cohen-tannoudji98,cool:phillips98},
providing friction and momentum diffusion in our system, and resulting
in a Brownian motion of the
atoms~\cite{cool:dalibard89,lat:sanchez-palencia02}.

In our setup, described in detail in
refs.~\cite{lat:ellmann03a,lat:ellmann03b,lat:petra06}, we superpose
two optical lattices, each being formed from the interference of four
laser beams, resulting in tetragonal lattice
structures~\cite{lat:grynberg01,lat:verkerk94}.  In lattice A, the
atomic state trapped is the $F_\mathrm{g}=3$ hyperfine structure
levels of the ground state of Cs (6s~$^2$S$_{1/2}$), while lattice B,
spatially overlapped with the former, traps atoms in the
$F_\mathrm{g}=4$ state.  The difference in energy between these two
ground states is large enough to enable spectrally selective optical
lattices, while at the same time being small enough that the
difference in the periodicity of the lattices (which depends on the
wavelength of the light) is negligible on the physical scale of the
region of interaction with the atoms.

The lasers are tuned near the D2 resonance (6s~$^2$S$_{1/2}$
$\rightarrow$ 6p~$^2$P$_{3/2}$) at 852 nm.  More specifically,
lattices A and B are tuned close to the $F_\mathrm{g}=3 \rightarrow
F_\mathrm{e}=4$ and $F_\mathrm{g}=4 \rightarrow F_\mathrm{e}=5$
resonances, respectively.  The latter is a closed transition, so the
rate of optical pumping out of lattice B will be slow, while the
former is an open transition, such that the probability of optical
pumping from lattice A to lattice B is high.  The relative spatial
phase between the two lattices can be controlled accurately, along any
direction in space, by changing the path lengths of the individual
laser beams~\cite{lat:ellmann03a,lat:ellmann03b,lat:petra06}.


The experimental sequence is as follows.  We start with a cloud of
laser-cooled Cs atoms at a temperature of a few microkelvins.  The two
optical lattices are then turned on, trapping some $10^8$ atoms, with
a filling fraction of about 5\%.  The atoms are left to interact with
the lattices for a time $\tau$, after which the lattice beams are
abruptly turned off.  Since no trapping potential is present any
longer, the atoms will fall due to gravity and reach a detection probe
located $\sim 5$~cm below the interaction region with the optical
lattices.  This time-of-flight technique~\cite{cool:phillips98} gives
the arrival time of atoms at the probe, which will depend on both the
vertical ($z$) position and the velocity of the atoms at the time of
release.  The initial velocity of the atoms in the double optical
lattice can then be straightforwardly extracted from this arrival
time, provided the distance between the interaction region and the
probe is known~\cite{lat:sjolund06}.

\begin{figure}
  \centerline{\includegraphics[width=0.45\textwidth]{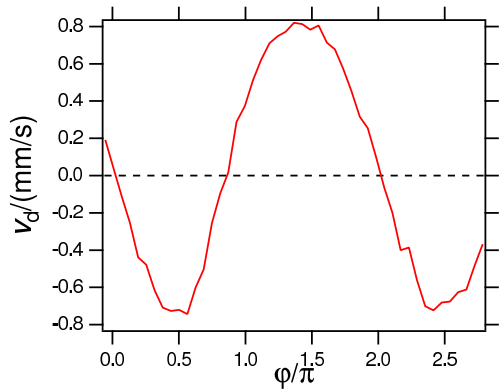}}
  \caption{Induced directed drift in the vertical ($z$) direction as a
    function of the relative spatial phase $\varphi$ for an
    interaction time $\tau$ of 350 ms.}
  \label{fig:tof}
\end{figure}

Sample data are presented in fig.~\ref{fig:tof},
where the velocity of the atomic cloud in the double optical lattice
is plotted as a function of the spatial shift between the lattices,
for lattice A detuned $35.9 \Gamma$ from the $F_\mathrm{g}=3
\rightarrow F_\mathrm{e}=4$ transition with an irradiance of
0.5~mW/cm$^2$ per laser beam, and lattice B detuned $40.0 \Gamma$ from
the $F_\mathrm{g}=4 \rightarrow F_\mathrm{e}=5$ transition, with
6.1~mW/cm$^2$ per beam ($\Gamma$ is the natural linewidth of the
6p~$^2$P$_{3/2}$ state).  As expected, no drift is observed when the
relative spatial phase, $\varphi$, is 0, $\pi$ or $2\pi$.  For all
other phase shifts, the data of fig.~\ref{fig:tof} clearly show an
induced drift, with opposite extrema around $\pi/2$ and $3\pi/2$.  The
maximum drift velocity observed here is $\approx 0.8\ \mathrm{mm/s}$,
or 1/4 of an atomic recoil.  It has been proven in
ref.~\cite{lat:sjolund06} that the drift velocity in the lattice is
constant in time, i.e., independent of the interaction time $\tau$.

As the lattice structures are periodic in all three dimensions and the
relative spatial shifts can be adjusted independently along the $x$,
$y$ or $z$ directions, the drift can be induced in any direction in
space.  This was evidenced in~\cite{lat:sjolund06} by direct imaging
of the atom cloud.

\section{Classical simulations}
\label{sec:simul}

In order to understand the qualitative behaviour of our Brownian
motor, we have performed simulations of a classical atomic cloud in an
optical lattice.  We thus consider Brownian particles which can be in
one of two internal states, indexed by $j$. The Brownian motion is
characterised by the momentum diffusion constant $\Dv(\vect{x})$ and
the external potentials $U_j(\vect{x})$.  The Fokker-Planck equation
(FPE)~\cite{risken89} for the probability distribution
$W_j(\vect{x},\vect{v},t)$ of a particle in state $j$ located at
$\vect{x}$ with velocity $\vect{v}$, written in time units of the
friction coefficient and in space units of the typical variation scale
of the potentials (so that all variables are dimensionless) is given
by~\cite{brown:sanchez-palencia04}
\begin{equation}
  \left[ \partial_t + v \partial_{\vect{x}} \right] W_j + \partial_{\vect{v}}
  \left[ \vect{v} + 
    \nabla\tilde{U}_j(\vect{x}) + \Dv(\vect{x}) \partial_{\vect{v}} \right] W_j 
  = \tilde{\gamma}_{j' \rightarrow j}(\vect{x}) W_{j'} -
  \tilde{\gamma}_{j \rightarrow j'}(\vect{x}) W_{j} .
\label{eq:FPE}
\end{equation}
The actual potentials created by the optical lattices are slightly
more complicated than the simple sine function model given in
fig.~\ref{fig:model}.  First, the different irradiances of the lasers
used to create each lattice results in distinct potential depths.
Moreover, the interaction strength also depends on the magnetic
sublevel of the atom, such that the latter will feel variations in the
potential as it is transferred from one lattice to the other and as it
is optically pumped between $M_F$ states.  Second, the actual
potentials in our system have different shapes in different
directions.

\begin{figure}
  \centerline{\includegraphics[width=0.45\textwidth]{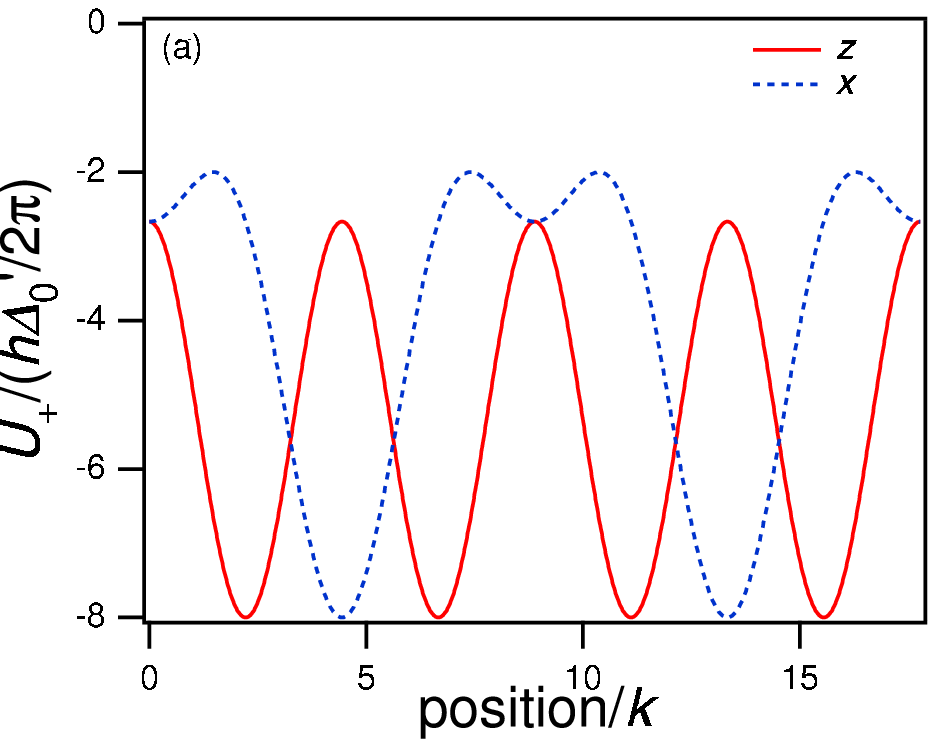} 
  \includegraphics[width=0.45\textwidth]{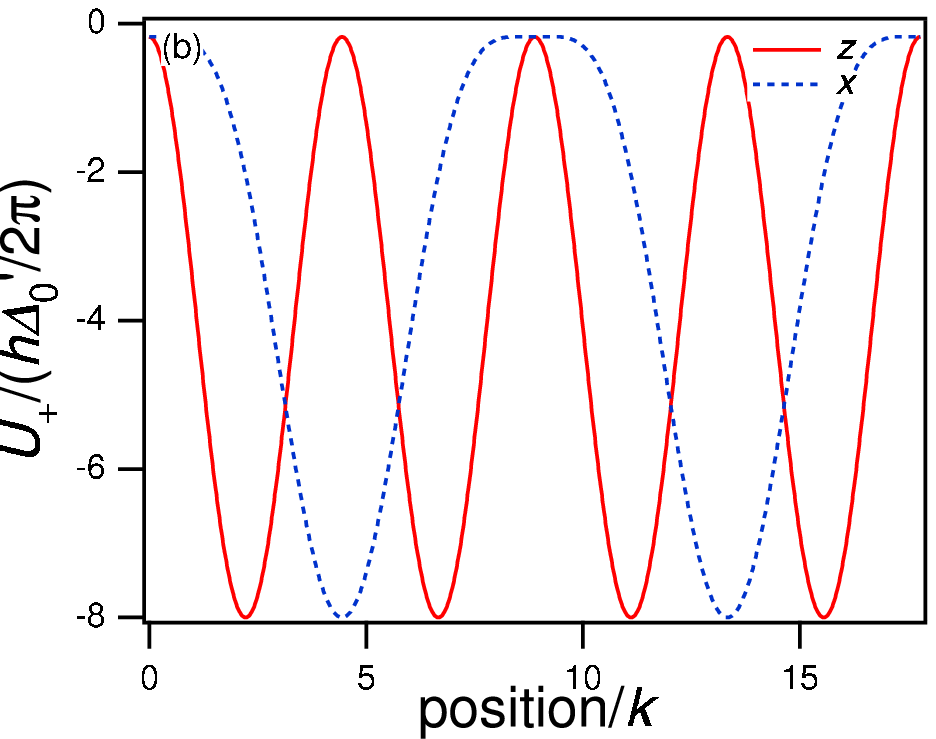}}
  \caption{Optical lattice potential along the $z$ [$U(x=0,y=0,z)$]
    and $x$ [$U(x,y=0,z=0)$] axes for (a) a $J_{\mathrm{g}} = 1/2
    \rightarrow J_{\mathrm{e}} = 3/2$ transition; (b) an atom in the
    $F=4$, $M_F = +4$ state, eq.~(\ref{eq:pot}).}
  \label{fig:pot}
\end{figure}

The simplest atomic system which undergoes a Sisyphus effect consists in
two atomic levels of total angular momentum $J_{\mathrm{g}} = 1/2$ in
the ground state and $J_{\mathrm{e}} = 3/2$ in the excited
state~\cite{cool:dalibard89}.  The resulting optical lattice potential
is then
\begin{equation}
  U_{\pm}(\vect{x}) = \frac{8 \hbar \Delta_0'}{3} \left[ \cos^2(k_xx) +
    \cos^2(k_yy) \mp \cos(k_xx) \cos(k_yy) \cos(k_zz) \right]
\label{eq:potH}
\end{equation}
for an atom in the $M_{\mathrm{g}} = \pm 1/2$ substate, where
$\Delta_0'$ is the light shift and the $k$'s are the effective
magnitude of the wave vectors along the axes~\cite{lat:grynberg01}.
We see that along $z$ the potential is simply sinusoidal, while in the
two transverse directions, deep wells are separated by shallower
minima, as shown in fig.~\ref{fig:pot}(a).

\begin{figure}
  \centerline{\includegraphics[width=0.45\textwidth]{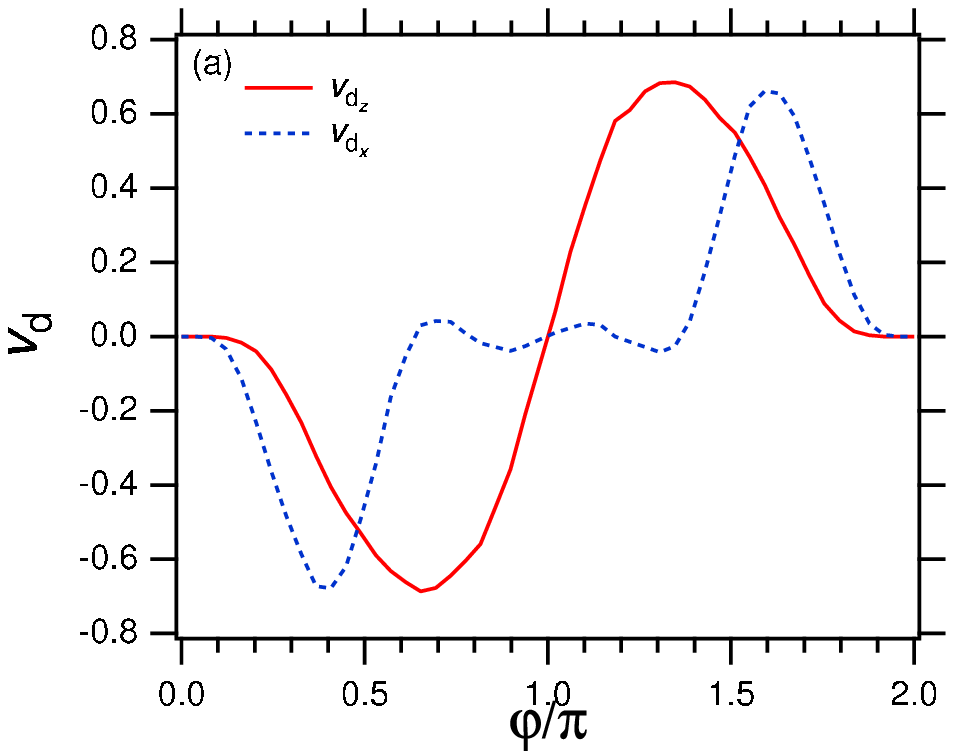}
    \includegraphics[width=0.45\textwidth]{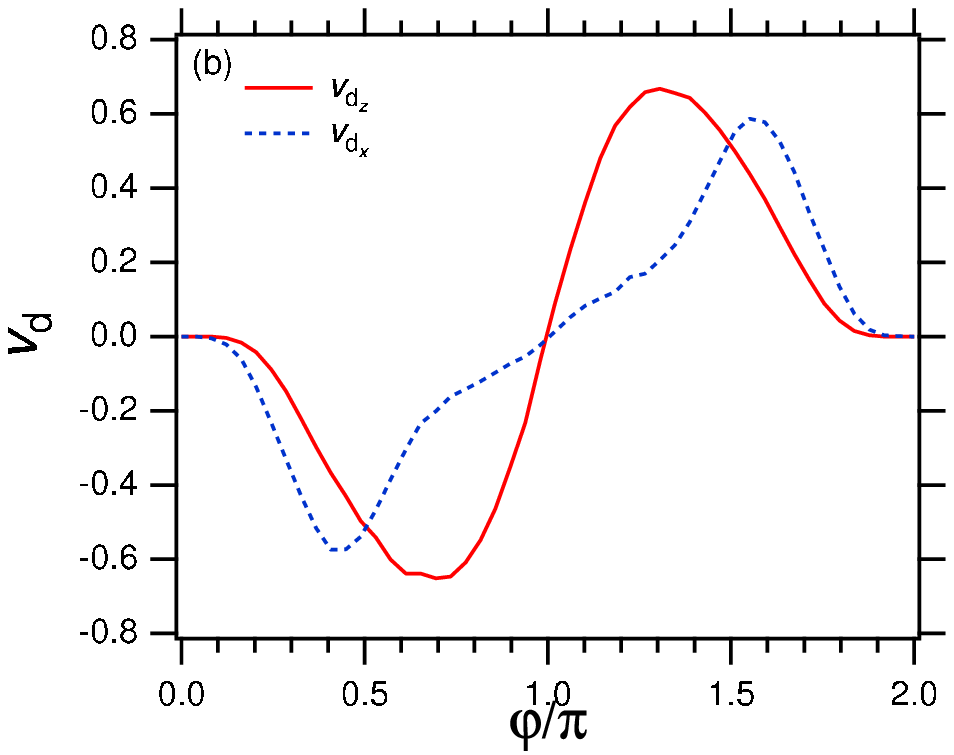}}
  \caption{Results from numerical simulations of the Brownian motor
    mechanism for the vertical ($z$) and horizontal ($x$) directions:
    (a) for a $J_{\mathrm{g}} = M_{\mathrm{g}} = 1/2$ atom; (b) for a
    $F_\mathrm{g} = M_{F} = 4$ atom. The drift velocity $v_\mathrm{d}$
    is plotted as a function of the relative spatial phase $\varphi
    \equiv \varphi_{\mathrm{A}} - \varphi_{\mathrm{B}}$, varied
    independently along $z$ or $x$.}
  \label{fig:simul}
\end{figure}

To get a qualitative picture of the effect of the shape of the
potential, we have performed numerical simulations of the
FPE~(\ref{eq:FPE}) in a 2D geometry, using the potential
\begin{equation}
  \tilde{U}_{j}(\vect{x}) = A_j \left[ \cos^2(x/2+\varphi_{j,x}) + 1 -
    \cos(x/2+\varphi_{j,x}) \cos(z+\varphi_{j,z}) \right],
\end{equation}
based on eq.~(\ref{eq:potH}) for $y=0$, with parameters $A = 200$,
$\Dv = 75$, and $\tilde{\gamma}_{\mathrm{A \rightarrow B}} = 3
\tilde{\gamma}_{\mathrm{B \rightarrow A}} = 7.5$, for a phase shift
chosen either in the $z$ or $x$ directions, resulting in a drift
velocity $v_{\mathrm{d}_z}$ or $v_{\mathrm{d}_x}$, respectively.
First, we see in fig.~\ref{fig:simul}(a)
that the overall variation of the drift velocity as a function of the
phase shift along $z$ is very similar to the one observed
experimentally (see fig.~\ref{fig:tof}).  One main difference is that
there is a plateau close to $\varphi = 0$ or $2\pi$ where the drift
velocity is very small, i.e., the Brownian motor mechanism is very
weak in these regions.  In contrast, in the experiment we get a clear
directed motion even for minute phase difference between the lattices.
Also, the simulation gives a maximum effect for phase shifts of
$\varphi \approx 2\pi/3$ and $4\pi/3$, while in the experiment the
maximum drift is seen at $\varphi \approx \pi/2$, $3\pi/2$.

We observe, as expected, that the shape of the potential has a strong
influence on the drift velocity for a given phase shift.  A
significant Brownian motor effect is seen only when the minima of the
deep potentials of both lattices are close together, but slightly
displaced.  This is easily explained as the atoms spend most of their
time close to the absolute minima of the long lived potential B, and
the force they feel when switching to the shorter-lived lattice A will
be bigger for the steeper slopes of the deep minima than the shallower
ones.

In reality, the Brownian motor with caesium atoms will show a less
pronounced difference along the different directions, as the hyperfine
structure is more complex than that obtained for the $1/2 \rightarrow
3/2$ atomic transition.  Optical pumping from the cooling process
leads to a polarisation of the atoms in the lattice, and most will be
found in the extreme $M_F = \pm F$ states.  For a caesium atom in the
$F_\mathrm{g} = 4$, $M_F = \pm 4$ state, the lattice potential is
given by
\begin{equation}
  U_{\pm}(\vect{x}) = \frac{4 \hbar \Delta_0'}{45} \left\{ 23 \left[
      \cos^2 \left( k_x 
        x \right) + \cos^2 \left( k_y y \right) \right] 
   \mp 44 \cos
    \left( k_x x \right) \cos \left( k_y y \right) \cos \left( k_z z
    \right) \right\}. 
  \label{eq:pot}
\end{equation}
The potential is still sinusoidal along $z$, while in the two
transverse directions, the minima are separated by regions where the
potential is essentially flat, as shown in fig.~\ref{fig:pot}(b).
Running the same simulations as before, but using now
\begin{equation}
  \tilde{U}_j(\vect{x}) = A_j \left\{ 23/44 \left[ \cos^2 (x/2+\varphi_{j,x}) +
      1 \right] 
   - \cos(x/2+\varphi_{j,x}) \cos(z+\varphi_{j,z}) \right\},
\end{equation}
we see in fig.~\ref{fig:simul}(b) that the difference between the $x$
and $z$ directions is less pronounced than before.

\section{Towards a quantum Brownian motor}
\label{sec:quantum}

As seen in the previous section, the main features of our Brownian
motor can be qualitatively described using a purely classical model.
This is because, although of an essential quantum nature, the coupling
between the potentials is driven by spontaneous emission processes
which can be treated as semi-classical random jumps~\cite{lindblad76}.
Nevertheless, it opens the way for the creation of a quantum Brownian
motor~\cite{brown:haenggi05b}. It could be realised at very low
temperatures, for instance, at high potential depths, where the motion
will clearly be quantised, or by going to a regime where tunnelling
between potential wells will become important. Another possibility is
to explore possible quantum resonances that are predicted for atomic
ratchets~\cite{lat:lundh05,brown:lundh06}.

One difficulty with the current setup is that the parameters that are
the potential depth, the magnitude of the diffusion, and the
transition rates between the two lattices all depend on both the
irradiance and the detuning of the optical lattice lasers.  Changing
either thus results in a different set of parameters.  Going to
far-detuned lattices and adding additional lasers to induce diffusion
and control the transfer between the lattices would allow for an
investigation of the effect of individual parameters.

\section{Conclusion}
\label{sec:conclusion}

In summary, we have realised a Brownian motor for cold atoms, based on
a dissipative double optical lattice. This is based on a new
rectification mechanism of noise in which the potentials are all
symmetric but spatially shifted with asymmetric transition rates.  As
our Brownian motor relies on spatially symmetric potentials, it allows
for a great control of the direction and magnitude of the induced
drift velocity.  Numerical simulations have shown that a simple model
describing the motion of Brownian particles shifting between two
sinusoidal potentials captures the main features of this Brownian
motor.  Because of the level of control it offers, our system is
promising for the study of the general properties and dynamics of
Brownian motors.  In particular, it should allow the investigation of
quantum versions of these.

\section*{Acknowledgements}

This research was conducted using the resources of the High
Performance Computing Center North (HPC2N).  We thank the Knut och
Alice Wallenbergs stiftelse, Carl Tryggers stiftelse,
Kempestiftelserna, Magnus Bergwalls stiftelse, the Swedish Research
Council, and SIDA/SAREC for financial support.


\end{document}